\documentclass[10pt,journal]{IEEEtran}

\usepackage{multicol}
\usepackage{lineno,hyperref}
\usepackage{amssymb}
\usepackage{float}
\usepackage{amsmath}
\usepackage{amsthm}
\usepackage{bbm}
\usepackage{graphicx}

\newtheorem{theorem}{Theorem}[] 

\begin{document}

\title{Secure symmetric ciphers over the real field}

\author{ 
\IEEEauthorblockN{Youssef Hassoun}

\IEEEauthorblockA{Department of Mathematics, American University of Science and Technology\\
                                Beirut, Lebanon\\
                                Email: youssef.hassoun@gmail.com
                               }
}

\maketitle
\begin{abstract}

Most cryptosystems are defined over finite algebraic structures where 
arithmetic operations are performed modulo natural numbers. 
This applies to private key as well as to public key ciphers.  
No secure cryptosystems defined over the field of real numbers 
are known. In this work, we demonstrate the feasibility of constructing 
secure symmetric key ciphers defined over the field of real numbers. 
We consider the security of ciphers introduced in a previous 
work and based on solving linear and non-linear equations numerically. 
We complement the design of those ciphers to satisfy the requirements 
of secure systems and, consequently, extend them into composite 
ciphers with multiple encryptions. We show security enhancements 
by estimating the uncertainty in finding the keys using a measure based 
on Shannon's entropy function.

\end{abstract}


\IEEEpeerreviewmaketitle

\section{Introduction}
\label{sec:intro}

Cryptography literature indicates that all known cryptosystems are 
defined over algebraic structures with finite sets. The order of these sets must 
be sufficiently large so as to make an exhaustive search (or brute-force attack) in 
the key space impractical. Both private key (symmetric) and public key (asymmetric) cryptosystems share this feature.  In addition, symmetric key cipher design must follow information theoretic principles to hide the internal structure of the encryption scheme, 
and, for public key ciphers, one-way trapdoor functions defined over finite structures 
are required; see for example,~\cite{menzes_1996, pieprzyk-2002, smart-2003} and~\cite{stinson-2005}, to name a few.
\newline

No cryptosystems defined over real numbers and implemented 
using floating point representations are known to be secure. Bergamo et al.~\cite{bergamo-2005} refer to examples of insecure symmetric key 
cryptosysyems based on chaotic maps. The authors also show that the 
public key cryptosystems of Kocarev and Tasev~\cite{kocarev-2003} 
based on chaos theory and use Chebyshev polynomials defined over 
real numbers are insecure. Due to the finite 
precision of the floating point representations of real numbers, rounding 
errors are unavoidable. For this reason, Kocarev and Tasev suggest upper 
bounds on numeric inputs to ensure the validity of 
the semi-group property of Chebychev polynomials, required to establish 
the public key protocol. On the other hand, finding a one-way trapdoor 
function over the field of real numbers similar to the function 
defined over the multiplicative group of a finite field, $f(x)=g^x\, (mod\, p)$, 
and known as the Discrete Logarithm Problem (DLP), seems to be difficult.
\newline

In symmetric key cryptography finding a one-way trapdoor function 
is not required and in case the design of encryption schemes allows 
rounding errors without loss of protocol correctness, it is possible
to construct secure symmetric key cryptosystems over real numbers. 
The purpose of the current work is to demonstrate the construction of
secure symmetric ciphers defined over the field of real numbers.
Cryptosystems based on real numeric root-finding methods have 
been introduced by the author in~\cite{hassoun-2014}. We address 
the security of these cryptosystems and extend them into 
multiple encryption product ciphers to meet the requirements of 
secure communications suggested by Shannon in his seminal work 
on communication theory of secrecy systems~\cite{shannon-1949}. 
To this end, we consider a security measure based on the entropy 
function proposed by Shannon in his earlier work on the theory of communication~\cite{shannon-1948}, and use this measure to 
estimate the security gained due to the addition of extra encryptions.
\newline

The contribution of this work can be summarized as follows:
\begin{itemize}
\item it extends the conference paper in :
\begin{itemize}
\item [i)] describing the proposed encryption schemes in more details   
\item [ii)] extending the encryption schemes into multiple encryption systems 
   (product ciphers) 
\item [iii)] addressing the security of the product ciphers against ciphertext only 
    attack and against known plaintext attack
\item [iv)] estimating the gained security by defining a security measure based on 
    Shannon's entropy function
\end{itemize}

\item it demonstrates the feasibility of constructing secure 
   symmetric key ciphers over the field of real numbers

\item it conjectures that a symmetric key cipher based on solving a 
   system of linear equations and defined over the field of real numbers is 
   secure against ciphertext only attack

\item it shows that security of symmetric key ciphers defined over the real field  depends solely on the size of key space, that is, on the precision and range of floating point representations of real numbers

\item it indicates that with appropriate software libraries, arbitrary precision is possible and, therefore, the key space size can be made arbitrarily large, thus allowing secure symmetric key ciphers over the field of real numbers  
\end{itemize}

This work is organized as follows: the next section~(\ref{sec:the_ciphers})
introduces the two private key ciphers based on numerical methods.  
Section~(\ref{sec:security}) addresses the security features of the proposed 
encryption schemes. In Section~(\ref{sec:prod}), the encryption schemes are 
extended to cascading cryptosystems involving multiple substitution and
transposition operations. Section~(\ref{sec:measure}) estimates the security 
gain due to the addition of extra encryptions, and, finally, Section~(\ref{sec:conc}) 
concludes the paper with a summary and future research directions.

\section{Cryptosystems defined over $\mathbbm{R}$} 
\label{sec:the_ciphers}
In this section, two encryption schemes defined over $\mathbbm{R}$
and based on employing numerical methods are described~\cite{hassoun-2014}.  
The first is a substitution $n$-block cipher ($n > 1$) based on solving 
a system of $n$ linear equations. The second is a one-character block 
cipher and is based on solving single non-linear equations numerically.

\subsection{Cipher based on solving a set of linear equations} 
The key consists of $n$-vector ($b_{i}$) and an $n\times n$ matrix
($a_{ij}$). Here, $n$ represents the block-length. Encrypting a block
of $n$ characters, represented by vector ($c_{i}$), is achieved by
solving the following system of linear equations:

\begin{equation}\label{eqn:lin_eqns} 
\sum_{j=1}^n a_{ij}x_j=b_i - c_i 
\end{equation}

Provided that $(a_{ij})$ is invertible, solution vectors ($x_{i}^*$)
exist; each represents a block of $n$ ciphertext entries.  The
condition on ($a_{ij}$) guarantees that the encryption function is
bijective and, consequently, has an inverse- the decryption function.
Decrypting the ciphered text is achieved by subsequently substituting
solution vectors into Equation~(\ref{eqn:lin_eqns}) giving rise to
$c_i=b_i-\sum_{j=1}^n a_{ij}x_j^*$.  One may re-write
Equation~\ref{eqn:lin_eqns} in an equivalent form:
 
\begin{equation}\label{eqn:lin_eqns_compact}
\sum_{k=1}^n  a_{ik}\times x_{kj}=b_i - c_{ij} 
\end{equation}

where $(x_{kj})$ represents the ciphertext; it is an ($n\times m$)
matrix with $m$ columns representing blocks, each of which is of size
$n$. The ($n\times m$) matrix $(c_{ij})$ represents the plaintext of
($nm$) characters; matrix $c_{ij}$ makes dividing the plaintext into
blocks, each of size $n$, explicit.
\newline

The following example serves to explain the encryption scheme by
comparing it to a similar classical substitution cipher, namely, the
Hill-cipher~\cite{hill-1929}. Consider a $2\times 2$ matrix key and a
constant vector defined as follows:
\begin{center} $A=\left( \begin{array}{ll} 2 & 3\\ 1 &
    4 \end{array} \right)$ and $\underline{b}=\left( \begin{array}{c}
    -3\\ 2 \end{array} \right)$ \end{center}.

Let ``epic" be part of a plaintext. The ascii code of this part can be
written in matrix form as $(c_{ij})=\left( \begin{array}{rr} 101 &
  105\\ 112 & 99\end{array} \right)$. The solution of the linear
  Equations~(\ref{eqn:lin_eqns_compact}) can then be written as
  $x_{ij}=(a^{-1})_{ik}\times(b_k-c_{kj})$, where $(a^{-1})_{ik}$ is
  the inverse matrix,
  $(a^{-1})_{ik}=\frac{1}{5}\left( \begin{array}{cc} 4 & -3\\ -1 &
    2 \end{array} \right)$, and $(x^*)_{ij}=\left( \begin{array}{ll}
    -17.2 & -28.2\\ -23.2 & -17 .2\end{array} \right)$ represents the
    ciphertext matrix. To decrypt and recover the plaintext matrix
    $(c_{ij})$, we substitute $(x^*)_{ij}$ in
    Equation~\ref{eqn:lin_eqns_compact}.
\newline

In implementing the Hill cipher the finite ring $\mathbbm{Z}_{26}$ 
is used, where all arithmetic operations are performed modulo
26. We remark that equivalent results will be obtained if the infinite 
real field $\mathbbm{R}$, employed in applying our algorithm in the 
previous paragraph, is used instead. To encrypt we calculate 
$a_{ik}\times c_{kj}\;(mod\;26) $ and obtain the ciphertext matrix
$y_{ij}=\left( \begin{array}{rr} 18 & 13\\ 3 & 7 \end{array}
\right)$. To decrypt, we calculate the inverse of $A\; (mod\;26)$,
$A^{-1}=\left( \begin{array}{ll} 6 & 15\\ 5 & 16 \end{array} \right)$,
and multiply by the ciphertext matrix modulo 26

\[ c_{ij}=(a^{-1})_{ik}\times y_{kj}=\left( \begin{array}{rl} 23 &1 \\ 8 &21 \end{array}
 \right)\equiv \left( \begin{array}{rr} 101 & 105\\ 112 &
   99\end{array} \right) \; (mod \; 26)\]

\subsection{Cipher based on solving non-linear equations}
\label{sec:solving_non_linear_eqns}
The key is a non-linear function with one variable $f(x)$. The encryption 
function is defined as finding a solution of the equation:
\begin{equation}\label{eqn:non-lin} f(x)-c_i=0   \end{equation} 
Here, $(c_i)$ represents the numerical code of the $i^{th}$ character
in the plaintext, e.g., the ascii-code. To guarantee that the
encryption function has an inverse, numerical encoding of plaintext
together with $f(x)$ must be chosen in such a way that
equation~(\ref{eqn:non-lin}) has at least one real root. The roots
\{$x_i^*$\} represent the ciphertext. On the recipient side, each
entry $(x_i^*)$ is decrypted by substituting it into $f(x)$, giving
rise to the plaintext character $c_i=f(x_i^*)$.  We remark that
\emph{{$f(x_i^*)$ must be appropriately rounded to recover $c_i$}}.
\newline

For example, given $f(x) = 2^{(x^2-x)}$ 
as key, encrypting ``epic'' amounts to solving, one at a time, 
four non-linear equations numerically: 
\begin{equation} \label{eqn:exp_fn}
  2^{(x^2-\frac{x}{2})}-c_i=0 
\end{equation}
where $c_i$ takes the values: 101, 112, 105 and 99. 
As a result, we get the following ciphertext using the Secant method:

\footnotesize
\[ \{x_i^*\}=\{-2.842433505.., 2.871040808.., 2.853218300.., 2.836862311..\} \] 
\normalsize

\noindent When those roots are substituted in~\ref{eqn:exp_fn}, we recover the \emph{real values}
of $c_i: 101.00..0, 112.00..0, 105.00..0\; and\; 99.00..0$ coinciding with the ascii code up to at least 
twelve decimal places.

\section{Security of the cryptosystems} 
\label{sec:security}
In this section, security features of the encryption schemes
introduced in the previous section are considered. Kerchoff's
Principle is assumed, that is, the specification of the encryption
algorithm is known but the key is unknown.
\newline 
    
There are two ways to break a private key cipher. One way is to try
all possible alternatives; the so-called exhaustive search or brute-force 
attack. This technique is guaranteed to succeed, but it is impractical if 
the key space is sufficiently large The second class of techniques is
based on making use of the internal structure of the cipher; for
example, as to how plaintext character blocks are mapped or encrypted
into ciphertext symbols. Such information helps the
adversary choose the most effective attack method.
\newline

Contemporary symmetric key ciphers, invented post 1970s, follow 
Shannon's two design principles of secure encryption~\cite{shannon-1949}:
\emph{confusion} and \emph{diffusion}. Contemporary ciphers are 
composite cryptosystems in which substitution is combined with 
transposition \emph{several times}; substitution adds confusion to the encryption 
process and transposition adds diffusion. Examples of design strategies 
underlying remarkable contemporary symmetric cipherse may explain 
how Shannon's design requirements are satisfied:
\begin{itemize}
\item[1)] Apply Feistel function (16 times) consisting of various operations, including expansion, key addition, S-Box substitution, P-Box permutation, and XOR, e.g., DES of the National Institute of Standards (US)~\cite{NIS-1977}
\item[2)] Employ three operations: modular addition, bit rotation and XOR,  e.g., RC5, suggested by R. L. Rivest~\cite{rivest-1994}
\item[3)] Mix two operations: a substitution and a permutation over finite fields, e.g., AES of the National Institute of Standards (US)~\cite{NIS-2001}
\end{itemize}

With the design principles of Shannon's information theory, plaintext
letters frequency distributions are diffused and, therefore, attacks
based on statistical analysis of ciphertexts, given the statistical
properties of the underlying language, fail. Diffusion breaks
monoalphabetic 1-to-1 correspondence between plaintext and ciphertext.
Next, we consider different attack models on the proposed
encryption schemes and discuss their robustness.

\subsection{Ciphertext only attack} \label{ssec:ciphertext_attack}
In this attack model, the adversary possesses only a copy of the
ciphertext.  Our cipher design based on solving a system of linear
equations, Equation~(\ref{eqn:lin_eqns_compact}), does not preserve
plaintext letters frequency distributions, it is a polyalphabetic
cryptosystem as the Hill cipher, and, therefore, statistical analysis 
using plaintext language redundancies is meaningless. According to 
Stinson~\cite{stinson-2005}, the Hill cipher is known to be hard to 
break in the ciphertext only attack model. The complexity of breaking 
the cipher can be evaluated by calculating the size of the key space. 
The size of the key space of Hill cipher with ($n\times n$) key matrix 
defined over finite ring $\mathbbm{Z}_{m}$ has been found by 
Overbey et al.~\cite{overbey-2005}:

\begin{theorem}\emph{(Overbey et al. Hill Cipher Keyspace Size Theorem)}\\
The number of ($n\times n$) matrices invertible mod $m=\prod_ip_i^{k_i}$ is
\[  |GL(n, \mathbbm{Z_m})| =\prod_i(p_i^{(k_i-1)n^{2}}\prod_{j=0}^{n-1}(p_i^n-p_i^j)) \]
\end{theorem}

Where $GL(n, \mathbbm{Z_m})$ represents the group of ($n\times n$) matrices 
invertible over $\mathbbm{Z_m}$. With $m=2\times 13$, the size of the key space 
will be:
\footnotesize{
\[  26^{n^2}(1-\frac{1}{2})\dots(1-\frac{1}{2^{n-1}})(1-\frac{1}{2^n})(1-\frac{1}{13})(1-\frac{1}{13^2})\dots(1-\frac{1}{13^n}) \]
}
\normalsize
 
An exhaustive search requires $26^{n^2}$ matrix multiplications.
Our cipher possesses a key space of a size bounded by the precision
of the floating point representation of real numbers. For example, the 
IEEE 754-1985 standard for binary floating point arithmetic~\cite{IEEE754-1985}, 
implemented by most compilers, and in particular the GCC compiler, defines an effective double-precision floating point range of 
$\pm(2-2^{-52})\times2^{1023}$ with $52$ bits of accuracy. 
This amounts to a range of $\approx 308$ decimal digits with 
$\approx 16$ decimal digits of accuracy. We remark that 
\emph{arbitrary precision} can be achieved 
using appropriate libraries, e.g., the MPFR library. 
Therefore, an exhaustive search can be made practically 
impossible. 
\newline

We conjecture that the cipher based on solving the system, 
Equation~(\ref{eqn:lin_eqns_compact}), defined over real numbers 
with double-precision floating point representation is secure in the 
ciphertext attack model.
\newline

The cipher based on solving non-linear equations numerically,
Equation~(\ref{eqn:non-lin}), is monoalphabetic and, therefore, 
is vulnerable against attacks based on statistical 
analysis using plaintext letters frequency distributions.

\subsection{Known plaintext attack}
In this attack model the adversary possesses a copy of (or a part) of
the plaintext as well as a copy of the corresponding ciphertext. In
the following, we denote plaintext characters by $\{c_i\}$ and their
corresponding ciphertext symbols by $\{x_i^*\}$.  Here, every $c_i$
corresponds to an ascii-code and every $x_i^*$ to a real number and
$i$ ranges over the size $l$ of (a part of) the plaintext.
\newline

The encryption scheme based on solving a system of linear
equations~(\ref{eqn:lin_eqns_compact}) is completely linear and,
therefore, vulnerable against known plaintext attacks. If the
dimension $(n)$ of the key matrix $(a_{ij})$ is known, a data set
consisting of an $(n+1)^2$ $(c_i,x_i^*)$ pairs is sufficient to
establish a system of linear equations whose solution is the key
matrix elements $(a_{ij})$ and $(b_i)$.
\newline

As indicated, the encryption scheme based on solving non-linear
equation~(\ref{eqn:non-lin}) is monoalphabetic; there is a 1-to-1 
correspondence from a subset of ascii-codes to a subset of 
$\mathbb{R}$ implying that the $(x_i^*)$ entries in data set 
$(x_i^*, c_i=f(x_i^*))$ are distinct for distinct $(c_i)$ entries, 
where $i\in [0..n]$ and $n$ represents the plaintext size.
With such a data set, it is possible to approximate the key function 
$f(x)$ to a polynomial function $p(x)$ of degree $\leq$ number  of 
distinct printable ascii characters. The existence of $p(x)$ is 
guaranteed by the following theorem due to Weierstrass~\cite{burden}:

\begin{theorem}
\emph{(Weierstrass approximation Theorem)}
\label{Weierstrass}
Suppose that $f(x)$ is defined and continuous on $[a, b]$. For 
each $\epsilon > 0$, there exists a polynomial $p(x)$, with the 
property that $|f (x) - p(x)| < \epsilon, for\; all\; x \in [a, b]$.
\end{theorem}

Given a data set, a unique polynomial $p(x)$ can be constructed using
interpolation. The construction procedure depends on the basis
polynomials of the vector space of dimension equals degree of $p(x)+1$.

\section{Applying Shannon's design principles} \label{sec:prod}
As mentioned in Section~\ref{sec:security}, contemporary symmetric 
ciphers are composite systems which combine a multiple of substitution 
and transposition operations in the encryption process to achieve reasonable
security levels. In the same section, we indicated that the proposed 
ciphers are restricted to substitution, and therefore, do not conform to 
Shannon's principles of secure  encryption. 
\newline

To improve security, we may extend the proposed encryption schemes into 
multiple encryption systems (product ciphers) by concatenating 
them to an arbitrary number of other block ciphers with independent keys. 
The following examples are restricted to extending the proposed schemes 
up to a maximum of three stages; they should serve as a proof of concept,
that is, to demonstrate the feasibility of constructing secure symmetric 
ciphers defined over the field of real numbers.

\subsection{Product cipher with solving linear equations}
\label{ssec:product_cipher_lin_eqns}
A 2-stage product cipher combining the proposed substitution cipher
based on solving systems of linear
equations~(\ref{eqn:lin_eqns_compact}) with an independent
transposition cipher would add the required \emph{diffusion} to the
encryption process. This would enhance the security against known 
plaintext attacks referred to in the previous paragraph. 
Symbolically, the product cipher encryption function can be
expressed as follows:

\begin{equation}\label{eqn:product_cipher_lin_enc}
  (e_{k_{tr}}\circ e_{k_{lin}}) (c_i)=e_{k_{tr}}(e_{k_{lin}} (c_i)))=x_i
\end{equation}  

The reverse operation (decryption) follows a reverse order and can be
symbolically expressed as:

\begin{equation}
 (d_{k_{lin}}\circ d_{k_{tr}}) (x_i)=d_{k_{lin}}(d_{k_{tr}} (x_i)))=c_i
\end{equation}
  
Here, $e_{k_{tr}}$ and $e_{k_{lin}}$ represent, respectively,
transposition encryption function and the encryption function based on
solving system of linear equations~(\ref{eqn:lin_eqns_compact}). The
symbols $d_{k_{tr}}$ and $d_{k_{lin}}$ represent, respectively, the
corresponding decryption functions, i.e., reversing transposition and
using the roots obtained to regain plaintext characters.
\newline

In implementing the first stage of the encryption process,
$e_{k_{lin}}$, the matrices $(a_{ij})$ and $(b_i)$ are read from a
text file and the root set $\{x_i^*\}$ was calculated by finding the
inverse matrix of equations~(\ref{eqn:lin_eqns_compact}) using the
formula
\[ A^{-1} = \frac{1}{det(A)} adj (A) \]
where $det(A)$ denotes the determinant of $A$ and $adj(A)$ is the adjoint matrix.
\newline 

\begin{figure}[H]
\centering 
\fbox{
\footnotesize{
\begin{minipage}{0.85\linewidth}
Input: ciphertext$_{in}$ generated by solving $n$ linear equations\\
Output: ciphertext$_{out}$\\ 
Let $f_p$ denote file pointer\\
$file\_size$
$\leftarrow$ calculate size of ciphertext$_{in}$\\ if ($file\_size$)
even, then $file\_size/2$ $\leftarrow$ $j$\\ else add a space char,
$(file\_size+1)/2$ $\leftarrow$ $j$;\\ $0$ $\leftarrow$ $i$; 0
$\leftarrow$ $k$; \\ 
while ($i < file\_size$) \&\& !eof(ciphertext$_{lin}$))
\begin{itemize}
 	\item[] read$(a_i)$ from input file; 
 	\item[] write$(a_i)$ in output file;
 	\item[] inc($k$); inc($i$);
 	\item[] if ($a_i==EOF$) then break;
 	\item[] else move $f_p$ to $j$, read$(a_j)$, write$(a_j)$,
          inc($j$), inc($i$), move $f_p$ to $k$;
\end{itemize}
end\\
end
\end{minipage}}}
\caption{Transposition algorithm}
\label{fig:transposition_algo}
\end{figure}

In transposing the encryptions generated by solving systems of linear
equations, we follow the algorithm shown in Figure~\ref{fig:transposition_algo}.
Figures~\ref{fig:champions_lin_enc} and~\ref{fig:champions_tr_enc}
show an example of the confusion and diffusion effects due to the two
encryption operations on part of a plaintext whose ascii-code is:
\begin{center} 
\fbox{
\footnotesize{
\begin{minipage}{0.55\linewidth}
087101032097114101032116104101032 \\ 
099104097109112115013010013010032
\end{minipage}}}
\end{center} 

Figure~\ref{fig:champions_lin_enc} depicts the ciphertext resulting
from the $1^{st}$ stage of the encryption process, that is, from
solving a set of ten linear equations
simultaneously. Figure~\ref{fig:champions_tr_enc} is the result of
applying the transposition algorithm on this ciphertext; the $2^{nd}$
stage of the encryption process. The matrix key employed in the first
stage is the following arbitrarily chosen ($10\times 10$) matrix ($a_{ij}$): 
\begin{center}
\fbox{
\footnotesize{
\begin{minipage}{0.96\linewidth}
\begin{tabular}{rrrrrrrrrr}
1&-1&-5&0.5&	-20&	0   &	0.4&	10&	0.25& 86\\
3&-1	 &0 &  2&-3&-12&52  & 1& 0&-0.1\\
0&23&9  &  9 & 3 &34 &-14& 7 & 9 &-8  \\
1&-9  &67 &-2 &-5 &8   & 20 &2&0.1&45 \\
-2 &23 &0&9 & 0 &34 &0.12&4    &3   &-4 \\
0.4&11 &1  &0  & 1  &0   &0.15&-0.8&89 &-1   \\
20 &0.2&-15&23 &-2   &1  &-10 &9  &23 &0.45 \\
0.5&-3 &0.1 &-30&-0.8&-3   &-12 &12&-11  &0.30 \\
-1 &-2 &	2  &	21&9    &-0.5&	35  &-3 &	-0.1&-1   \\
3  &0  &-1   &-0.1&11  &0    &-2   &7  &9     &0.8  
\end{tabular}
\end{minipage}}}
\end{center}

The constant $b_j$ vector is chosen to be:
\begin{center}
\fbox{
\footnotesize{
$(-10\; 2\; 27\; -1\; 90\; 0.2\; -4\; 12\; 30\; -0.5)^T$ 
}}
\end{center}

\begin{figure}[h]
\centering 
\fbox{
\scriptsize{
\begin{minipage}{1.0\linewidth}
-9343.900391 -1072.250000 -6781.200195 -5534.299805 -6628.520020
-7563.500000\\ -6515.274414 3477.149902 -2777.700195 -1943.399902
-442.599976 -5014.049805\\ -5717.200195 -7918.899902 -6734.479980
596.650024 -397.275085 4744.850098\\ -6241.600098 152.000000
\end{minipage}}}
\caption{Ciphertext from $1^{st}$ stage (solving lin eqns) of
  encryption process}
\label{fig:champions_lin_enc}
\end{figure}

\begin{figure}[h]
\centering 
\fbox{
\scriptsize{
\begin{minipage}{0.98\linewidth}
-33909 17.500-71209 53.985-68502 76.000-55241 47190 27.015-93390 42597
51.485\\-77209 71.992-74498 9.504-9.70544.508-21609 5.00094.031-02200
68.015-54290 62.200-53500 61.74437.492-77709 14.992-4.996-04090
51.015-98890 63.79056602 37258 74809 64.00812000
\end{minipage}}}
\caption{Ciphertext from $2^{nd}$ stage (transposition) of encryption process}
\label{fig:champions_tr_enc}
\end{figure}

\begin{figure}[h]
\centering 
\fbox{
\footnotesize{
\begin{minipage}{0.8\linewidth}
Input: ciphertext$_{tr}$ due to transposition cipher\\ 
Output: ciphertext$_{lin}$ \\ 
Let $f_p$ denote file pointer\\
$file\_size$ $\leftarrow$
calculate size of ciphertext$_{tr}$ \\ 
 if ($file\_size$) even then $file\_size/2$ $\leftarrow$ $j$\\ 
 else add a space char, $(file\_size+1)/2$ $\leftarrow$ $j$;\\
$0$ $\leftarrow$ $i$;
$j$ $\leftarrow$$l$\\ 
while ($i < l$ \&\& $j < file\_size$)
\begin{itemize}
 	\item[] read $a_i$;  /* $a_i$ denotes a ciphertext$_{tr}$ element */
 	\item[] write $a_i$ in ciphertext$_{lin}$;
 	\item[] inc($i$); 
 	\item[] move $f_p$ to position $j$; 
 	\item[] read $a_j$; write $a_j$ in ciphertext$_{lin}$; inc($j$);
 	\item[] move $f_p$ to position $i$; 
\end{itemize}
end\\
	if ($i < l$) then read $a_i$; write $a_i$ in ciphertext$_{lin}$;\\
end
\end{minipage}}}
\caption{Inverse transposition algorithm}
\label{fig:inv_transposition_algo}
\end{figure}

The inverse decryption process proceeds with reversing the
transposition process $(d_{k_{tr}} (x_i))$, that is, applying the
inverse transposition algorithm of
Figure~\ref{fig:inv_transposition_algo}. This operation recovers the
ciphertext of roots generated by solving the system of linear
equation~\ref{eqn:lin_eqns_compact}. The roots are then substituted
into the linear equations thus recovering the plaintext characters
$c_i$.  Figure~\ref{fig:champions_lin_dec} shows the result of the
$2^{nd}$ stage of decrypting process, i.e., inserting the roots of the
ciphertext into the linear equations to get the ascii-code. The
resulting real numbers must be rounded to recover the plaintext.

\begin{figure}[H]
\centering 
\fbox{
\scriptsize{
\begin{minipage}{0.95\linewidth}
87.000114 101.000092 32.000019 97.000038 113.999611 100.999565
31.999895\\ 116.000237 104.000084 100.999886 32.000282 99.000137
103.999985 97.000183\\ 108.999832 111.999611 114.999886 13.000096
10.000035 12.999951
\end{minipage}}}
\caption{Ciphertext from $2^{nd}$ stage of decryption process}
\label{fig:champions_lin_dec}
\end{figure}

\begin{figure}[H]
\centering
\includegraphics[width=0.50\textwidth]{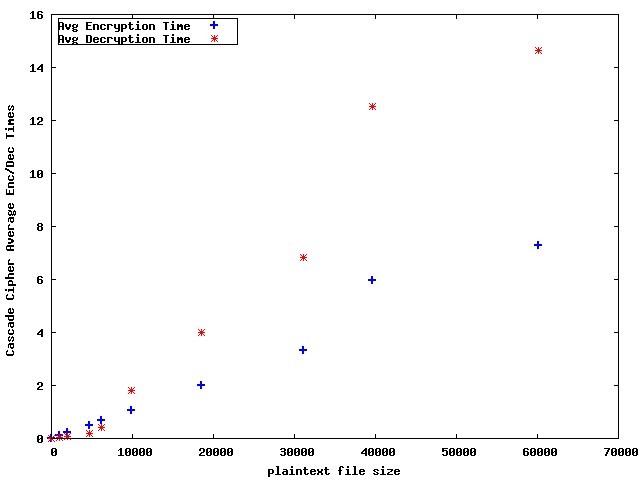}
\caption{Average enc/dec times of product cipher}
\label{fig:lin_tr_3x3}
\end{figure}

A correlation analysis of the encryption/decryption time versus the
size of plaintext of the product cryptosystem combining solving
systems of linear equations and transposition reveals a linear model
with correlation coefficients: $r_{enc}=0.9876$ and $r_{dec}=
0.9832$. Figure~\ref{fig:lin_tr_3x3} exhibits the linear correlations.
The sizes of plaintext samples were: 21, 1036, 2024, 4658, 6218, 9830,
18552, 31081, 39674, 60173 bytes and the matrix key was arbitrarily
taken to be a $3\times 3$ matrix.

\subsection{Product cipher with solving non-linear equations}
\label{ssec:product_cipher_non_lin_eqns}
In the following, a 3-stage product cipher resulting from
concatenating our encryption function based on solving
equation~(\ref{eqn:non-lin}) with the Vigen\`ere cipher and with a
transposition cipher is constructed. Adding the Vigen\`ere and
transposition operations breaks the ``monoalphabetic" link between
ciphertext and plaintext (the input) by spreading out plaintext
redundancy over entire ciphertext.
\newline

The triple encryption cryptosystem is a product cipher whose
encryption function can be symbolically expressed as follows:

\begin{equation} \label{eqn:product_cipher_non_lin_enc}
 (e_{k_{tr}}\circ e_{k_{vig}} \circ e_{k_{nlin}}) (c_i)=e_{k_{tr}}(e_{k_{vig}}(e_{k_{nlin}} (c_i)))=x_i
\end{equation}
  
\noindent where $e_{k_{tr}}$ represents the transposition function,
$e_{k_{vig}}$ represents the Vigen\`ere encryption function with
$k_vig$ as keyword, $e_{k_{bis}}$ represents the encryption function
based on solving equation~(\ref{eqn:non-lin}) with $k_{nlin}$
representing the non-linear function $f(x)$, and $c_i$ is the $i^{th}$
plaintext character with $x_i$ being the corresponding ciphered
character. The inverse decryption function follows a reverse order,
namely,
\begin{equation}
 (d_{k_{nlin}}\circ d_{k_{vig}}\circ d_{k_{tr}}) (x_i)=d_{k_{nlin}}(d_{k_{vig}}(d_{k_{tr}} (x_i)))=c_i
\end{equation}

\noindent where $d_{k_{tr}} (x_i)$ represents the transposition
inverse, $d_{k_{vig}} (x_i)$ represents the Vigen\`ere decryption
function which recovers the root $x^*_i$ generated by solving the
corresponding non-linear equation with key function $k_{nlin}$.
$d_{k_{nlin}}(x_i)$ represents the decryption function, by which the
root $x_i$ is substituted into the non-linear equation to recover the
plaintext character $c_i$.
\newline

We apply the three encryption operations on the same (part of)
plaintext with the ascii-code given in the previous
section. Figure~\ref{fig:non_lin_enc} shows the ciphertext resulting
from the first encryption operation ($e_{k_{nlin}}$) based on solving
the following non-linear equation numerically using the Bisection
method:
  \[ x^5+7.34x^4+22.03x^3+46.012x^2+12.25x-1-c_i  = 0\]

\begin{figure}[h]
\centering 
\fbox{
\scriptsize{
\begin{minipage}{0.98\linewidth}
0.996905152715 1.062095760863 0.632444388903 1.044151171664
1.117163734307\\ 1.062095760863 0.632444388903 1.125235020154
1.075229083508 1.062095760863\\ 0.632444388903 1.053187075449
1.075229083508 1.044151171664 1.096536606346\\ 1.108991331275
1.121211836726 0.402103486558 0.350298562407 0.402103486558
\end{minipage}}}
\caption{Ciphertext from $1^{st}$ stage (solving non-lin eqn) of encryption process}
\label{fig:non_lin_enc}
\end{figure}

The Vigen\`ere encryption keyword ($k_{vig}$) takes the form of an
array of real numbers of a certain length, $keyword[]$. The Vigen\`ere
cipher adds, in order, each root arising from the first encryption
operation (Figure~\ref{fig:non_lin_enc}) to an array element in the
keyword, $y_i=x^*_i+keyword[i]$. Depending on the length of the
keyword compared to the length of the ciphertext, the same roots will
be, with high probability, assigned different encryption values, thus
breaking the 1-to-1 correspondence between plaintext letters and
ciphertext real roots. Figure~\ref{fig:vig_keyword} depicts the 10
entries of Vigen\`ere's keyword used in encrypting the
roots. Figure~\ref{fig:vig_ciphertext} lists the ciphertext elements
generated by applying the Vigen\`ere cipher to the roots using the
keyword of Figure~\ref{fig:vig_keyword}. We remark that the 1-to-1 
link has been eliminated; equal roots do not anymore correspond to 
equal ciphertext elements.

\begin{figure}[H]
\centering 
\fbox{
\scriptsize{
\begin{minipage}{0.93\linewidth}
8.27409124359 3.44876404589 2.84907100186 1.27800971542
4.90898111008\\ 5.46406511234 0.21409875231 7.19061419871
2.38408754321 3.12908182363
\end{minipage}}}
\caption{Vigen\`ere's cipher keyword of length 10} 
\label{fig:vig_keyword}
\end{figure}

\begin{figure}[H]
\centering 
\fbox{
\scriptsize{
\begin{minipage}{0.98\linewidth}
9.270995914936 4.510859847069 3.481515407562 2.322160959244
6.026145100594\\ 6.526160836220 0.846543133259 8.315849184990
3.459316611290 4.191177487373\\ 8.906535148621 4.501951217651
3.924300074577 2.322160959244 6.005517959595\\ 6.573056459427
1.335310637951 7.592717707157 2.734386116266 3.531185209751
\end{minipage}}}
\caption{The ciphertext generated after applying the Vigen\`ere
  cipher}
\label{fig:vig_ciphertext}
\end{figure}

The third and last encryption operation due to transposition
($e_{k_{tr}}$) is applied to the ciphertext of
Figure~\ref{fig:vig_ciphertext} as input using an implementation of
the algorithm depicted in
Figure~\ref{fig:transposition_algo}. Figure~\ref{fig:tr_ciphertext}
depicts the ciphertext.

\begin{figure}[H]
\centering 
\fbox{
\scriptsize{
\begin{minipage}{0.94\linewidth}
9209943 .1894093411476 .2105246064109 .2103200864135 .1898903491619
\\.9178338-963182 .0911613940047 .2105246051999 .7065471351675\\
.9770172748166 .315071-.7951964505876 .8550522326994 .2150546566862\\
.4533298354149 .5361204117477 .0554614515275 .2307572326994\\
.0575556535492 .3303917521775 .3361263518295

\end{minipage}}}
\caption{The ciphertext generated after applying the transposition cipher}
\label{fig:tr_ciphertext}
\end{figure}

The inverse decryption process proceeds with reversing the
transposition process $(d_{k_{tr}} (x_i))$) applied to the last
ciphertext generated by the transposition
operation-Figure~\ref{fig:tr_ciphertext} as input. This operation must
result in the ciphertext of Figure~\ref{fig:vig_ciphertext}, i.e.,
that generated by applying Vigen\`ere's cipher in the encryption
process. To decipher this text we subtract, in order, each real value
($y_i$) from the corresponding array element of Vigen\`ere's keyword
of Figure~\ref{fig:vig_keyword}. In this way the roots arising from
the first encryption operation are recovered, $x^*_i=y_i -
keyword[i]$. The roots are then substituted in the polynomial function
to recover the plaintext characters $c_i$.
\newline

\begin{figure}[H]
\centering
\includegraphics[width=0.50\textwidth]{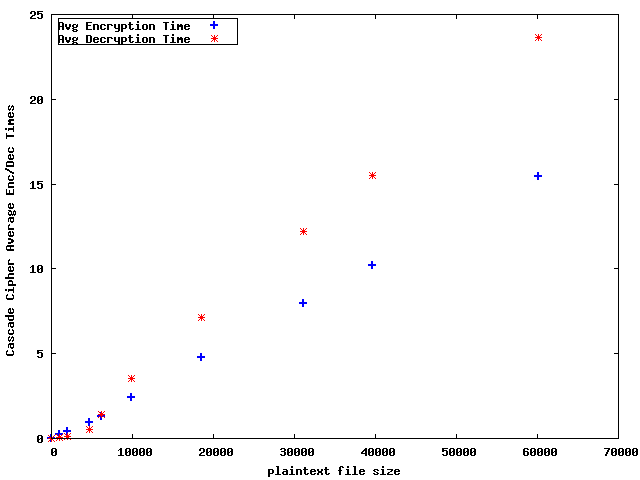}
\caption{Average enc/dec times of product cipher}
\label{fig:bis_vig_tr_times}
\end{figure}

Following the correlation analysis of data performed in
Section~\ref{ssec:product_cipher_lin_eqns}, a similar analysis of the
encryption/decryption times versus the size of plaintext of the
product cryptosystem combining solving non-linear single equations,
Vigen\`ere cipher and transposition operation reveals a linear
correlation with coefficients: $r_{enc}=0.9998$ and
$r_{dec}=0.9988$ (See Figure~\ref{fig:bis_vig_tr_times}). 
The linear correlation of the cipher based on
solving non-linear equations numerically
persists in the product cipher with greater response times. This behavior 
shows that the encryption/decryption times due to Vigen\`ere encipherment
together with the transposition cipher, increase linearly with the size of 
plaintext.

\section{Estimating the gained security} \label{sec:measure}
In this section we use Shannon's probabilistic approach to estimate
the security gained by extending our cryptosystems into product ciphers
against ciphertext only attacks. We start with a brief introduction to
the basic concepts of this approach required to define a security measure.
\newline

A cryptosystem is represented as 5-tuple $\mathcal{(P, C, K}, e_K(), d_K())$,
where $\mathcal{P}$ represents the set of possible plaintexts 
(single letters of the alphabet of the underlying plaintext language), 
$\mathcal{C}$ the set possible ciphertexts, $\mathcal{K}$ the set of 
possible keys, $e_K()$ the set of possible encryption functions with 
$K$ being a random variable which takes values in  $\mathcal{K}$
according to a (possibly uniform) probability distribution, and
$d_K()$ represents the corresponding decryption functions. 
\newline

Shannon~\cite{shannon-1948} defined the entropy function as a measure 
of uncertainty (in bit length) in the following way. Given a random variable 
$X$ with probability distribution $\mathit{p}(X)$, the entropy $H(X)$ is:

\begin{equation}\label{eqn:entropy} 
 H(X) =  - \sum_{i=1}^n{\mathit{p}(X=x_i)\log_2(\mathit{p}(X=x_i))} 
\end{equation}
Here, $x_i$ represents all possible values of $X$ and (finite) $n$ being the
size of the sample space on which $X$ is defined. We remark that 
the maximum value ($H(X)=\log_2(n)$) is reached in case $\mathit{p}(X)$ 
is uniform, i.e., $\mathit{p}(X=x_i)=\frac{1}{n}$ for all $i$, and the minimum
value ($H(X)=0$) in case one single X-value is certain, i.e., 
$\mathit{p}(X=x_{i_0})=1$ and $\mathit{p}(X=x_i)=0$ for all $i\neq i_0$.
\newline

Key equivocation will be used as a measure of security; it is a conditional 
entropy expression of the form $H(K|C)$ and measures the average 
uncertainty remaining about the key when a ciphertext has been 
observed. One can show that, following Stinson~\cite{stinson-2005}:

\begin{equation}\label{eqn:key_equivoc} 
H(K|C^n)=H(K) + H(P^n) - H(C^n)
\end{equation}
where $P^n$ represents the random variable that has as its probability
distribution all $n$-grams of plaintext alphabet, $C^n$ is a random variable 
with a probability distribution being all $n$-grams of ciphertext symbols and 
$K$ the random variable with a uniform probability distribution. 
\newline

In case $\mathcal{P}$ represents the alphabet of a natural language (L),
one can show that $H(P^n) \approx nH_L$, where $H_L$ represents the 
entropy of a single letter. For English, $H_{E}\approx 1.25$. Also, %
as the upper bound on entropy $H(X)$ of a random variable $X$ taking
$l$ values $x_1, x_2, \dots, x_l$ with \emph{any} probability
distribution $\mathit{p}(X)$ is $\log_2(l)$, we have
$H(C^n)\le n\, \log_2(|\mathcal{C}|)$. Putting these results
in~(\ref{eqn:key_equivoc}), we get, following Stinson~\cite{stinson-2005}:

\begin{equation}\label{eqn:key_equivoc_fin} 
H(K|C^n) \geq H(K) + nH_L - n\, \log_2(|\mathcal{C}|)
\end{equation}

In case $\mathcal{P}$ does not correspond to a natural language, 
the language redundancy is zero, e.g., the set of digits, and if 
$|\mathcal{P}|=|\mathcal{C}|$ and we assume that the probability 
distributions of $P^n$ and of $C^n$ are the same, then 
$H(P^n) - H(C^n)=0$ and, therefore, equation~\ref{eqn:key_equivoc} 
reduces to 
\begin{equation} \label{eqn:key_equiv_spec}
H(K|C^n)=H(K)
\end{equation}

Equation~(\ref{eqn:key_equiv_spec}) tells us that in case the plaintext 
language has no redundancy, that is, all its symbols are equally probable, 
there is no information about the key that can be conveyed by the ciphertext 
$n$-gram, and, therefore, the uncertainty about the key is entirely 
determined by the size of the key space. Next we use this result to estimate
the security gained by adding Vigen\`ere and transposition encryptions.

\subsection{Product cipher with solving linear equations}
As indicated in Section~\ref{ssec:ciphertext_attack}, the size of the 
space of our cryptosystem based on solving a set of linear equations 
simultaneously is infinite; there are infinite number of invertible square 
matrices over $\mathbbm{R}$. This makes exhaustive search much 
harder than the case of Hill cipher defined over a finite field. 
\newline

As for the additional uncertainty in finding the key added by applying the 
transposition cipher, we have $H(K_{tr}|C^n)=H(K_{tr})$, since the 
transposition is a mapping from $\mathbbm{R}$ into $\mathbbm{R}$. 
If block length $n$ is even and we apply the transposition encryption 
algorithm Figure~\ref{fig:transposition_algo}, the size of key space is 
$|\mathcal{K}_{tr}|=(\frac{n}{2})\,!$ Therefore, 
$H(K_{tr})=\log_2(\frac{n}{2})\,!\approx \frac{n}{2}\log_2(\frac{n}{2e})+\log_2\sqrt{\pi n}$ 
is the additional key uncertainty gained by applying transposition. For $n=100$, the 
gained uncertainty is $\approx 214.2$ bits.

\subsection{Product cipher with solving non-linear equations}
The cryptosystem of Section~\ref{ssec:product_cipher_non_lin_eqns} consists
of three encryption stages. We start with the last stage, where transposition 
encryption is applied on the ciphertext, a string of real numbers, produced from 
the second stage, the Vigen\`ere encryption. As indicated in the previous section,
with $n$ being (even) length of the ciphertext due to Vigen\`ere's encryption.
The additional uncertainty in finding the key is:
$H(K_{tr})=\log_2(\frac{n}{2})\,!\approx \frac{n}{2}\log_2(\frac{n}{2e})+\log_2\sqrt{\pi n}$
\newline

At the second stage the Vigen\`ere encryption is applied to the ciphertext 
consisting of the real roots of non-linear equations; it is a mapping
from $\mathbbm{R}$ into $\mathbbm{R}$, in which case $H(K|C^n)=H(K)$.
Let the length of the keyword- a string of digits, be $k$. The size of the key space 
will be $|\mathcal{K}_{vig}|=10^k$. Therefore, $H(K_{vig})=\log_2(10^k)=k\log_2(10)$ 
\newline

At the first stage the encryption is monoalphabetic and we discard its 
contribution to the overall security of the cryptosystem. The key space of the 
third and the second encryptions is the product of both spaces, and, therefore,
the gained uncertainty of the system is the sum 
\[ H(K_{tr})+H(K_{vig})\approx \frac{n}{2}\log_2(\frac{n}{2e})+\log_2\sqrt{\pi n}+k\log_2(10) \]

For $n=100$ and k=$20$, the gained uncertainty amounts to $\approx 217.5$ bits.

\section{Conclusions} \label{sec:conc}
We demonstrated that it is possible to build secure symmetric key
cryptosystems over the field of real numbers. The starting
point was encryption schemes based on root-finding numerical methods. 
The designs of these schemes is restricted to substitution, and, therefore, 
do not satisfy Shannon's principles of secure communication.
\newline

We applied Shannon's security principles and extended the proposed schemes into 
multiple encryption product ciphers. With this extension, the resulting ciphers 
become similar to contemporary symmetric ciphers such as DES and AES,
with the difference that the latter ciphers are defined over finite fields, whereas 
ours are defined over the field of real numbers. However, implementing the 
ciphers on a computing machine with finite memory sets upper and lower 
bounds on the size of the real field; these bounds are determined by the 
range and the precision of floating point representations of real numbers. 
\newline

A security measure based on Shannon's entropy function is used to estimate
security against ciphertext only attacks gained by adding more encryptions, 
like Vigen\`ere and transposition operations. As parts of the multi-stage encryption 
process, these encryptions are mappings from the set of reals $\mathbbm{R}$ into 
itself. With such mappings, there is no redundancy in the source (plaintext) 
language and the probability distributions of source and ciphertext languages are 
equal. As a result, the uncertainty of finding the corresponding keys depends 
only on the size of the key space, which, as noted, depends on the range 
and precision of floating point representations of real numbers. We indicated 
that arbitrary precision can be achieved using appropriate software libraries and, therefore, the key space size can be made arbitrarily large. This meant
that exhaustive search in the key space can be made practically impossible, thus enhancing the security of our product ciphers.
\newline

Following the demonstrations presented in this work, future research will concentrate 
on building secure large scale (industrial) symmetric key cryptosystems defined over the field of real numbers.   

\section*{Acknowledgment}
The author would like to thank Dr Steve Counsell of Brunel University, UK, and Hiba Othman of the American University of Science \& Technology, Beirut, for critical reading of the manuscript.

\end{document}